\documentclass{SCGE}
\usepackage{multirow}
\usepackage{hyperref}
\usepackage{url}

\begin{document}

\begin{picture}(0,0){\rm
\put(0,-20){\makebox[160truemm][l]{\bf {\sanhao\raisebox{2pt}{.}}
Review Article  {\sanhao\raisebox{1.5pt}{.}}}}}
\put(0,-34){\jiuwuhao {\textcolor[rgb]{0.5,0.5,0.5}{\sf 
}}}
\end{picture}

\def\bm{\boldsymbol}

\def\dl{\displaystyle}
\def\du{\end{document}}
\def\d{{\rm d}}
\def\e{{\rm e}}
\def\i{{\rm i}}
\def\pi{{\uppi}}

\Year{2015} %
\Month{Nov} %
\Vol{00} 
\No{00} 
\BeginPage{00} 
\AuthorMark{{\rm Zhu et al.}}  
\AuthorMarkCite{{\rm Zhu X.-J. et al. }} 
\DOI{0000} 
\ArtNo{0000}

\title[Detecting nanohertz gravitational waves with pulsar timing arrays]
{Detecting nanohertz gravitational waves with pulsar timing arrays}

\author[1,2*]{Xingjiang ZHU}{}
\author[1*]{Linqing WEN}{}
\author[2]{George HOBBS}{}
\author[2]{Richard N. MANCHESTER}{}
\author[2,3]{Ryan M. SHANNON}{}

\address[{\rm1}]{School of Physics, University of Western Australia, Crawley WA 6009, Australia;}
\address[{\rm2}]{CSIRO Astronomy and Space Science, PO Box 76, Epping NSW 1710, Australia;}
\address[{\rm3}]{International Centre for Radio Astronomy Research, Curtin University, Bentley, WA 6102, Australia}

\maketitle \vspace{-3.5mm}{\footnotesize\begin{center} Received xxx, 2015; accepted xxx, 2015; published online xxx, 2015
\end{center}}\vspace*{-5mm}

\begin{center}
\rule{16.5cm}{0.4pt}
\parbox{16.5cm}
{\begin{abstract} Complementary to ground-based laser interferometers, pulsar timing array experiments are being carried out to search for nanohertz gravitational waves. Using the world's most powerful radio telescopes, three major international collaborations have collected $\sim$10-year high precision timing data for tens of millisecond pulsars. In this paper we give an overview on pulsar timing experiments, gravitational wave detection in the nanohertz regime, and recent results obtained by various timing array projects.
\end{abstract}}
\end{center}\vspace*{-0.6cm}

\begin{center}
\parbox{16.5cm}
{\bf\jiuhao pulsars, gravitational waves, black-hole binaries}
\end{center}

\begin{center}
{}
\CITA    
\end{center}

\textwidth=178truemm \textheight=236truemm

\wuhao\vspace*{1.5mm}

\begin{multicols}{2}

\renewcommand{\baselinestretch}{1.08} \baselineskip 12.2pt\parindent=10.8pt

\section{Introduction}\label{sec:intro}

One hundred years ago, Albert Einstein completed his general theory of relativity, radically revolutionizing our understanding of gravity. In this theory, gravity is no longer a force, but instead an effect of spacetime curvature. The mass and energy content of spacetime creates curvature that in turn dictates the behavior of objects in spacetime. Generally speaking, when a massive object is accelerating, it produces curvature perturbations that propagate at the speed of light. Such \emph{ripples} in the fabric of spacetime are called gravitational waves (GWs).

Forty years ago, Hulse and Taylor discovered the binary-pulsar system PSR B1913+16 \cite{HulseTaylor75}. Subsequent observations in the following years showed that its orbital period was gradually decreasing, at a rate that was entirely consistent with that predicted by general relativity as a result of gravitational radiation \cite{Taylor82,Weisberg10}. This provided the most convincing evidence of the existence of GWs and earned Hulse and Taylor the Nobel prize in Physics in 1993.

\vspace{-1mm}
\noindent\rule{2.5cm}{0.4pt}\\\vspace{-1.5mm} {\qihao *Corresponding author (Xingjiang ZHU, email: xingjiang.zhu@uwa.edu.au; Linqing \vspace{0.3mm}WEN, email: linqing.wen@uwa.edu.au)}

It was realized in late 1970s that precision timing observations of pulsars can be used to detect very low frequency GWs ($\sim 10^{-9}$--$10^{-7}$ Hz; \cite{Sazhin1978,Detweiler1979}). Based on a calculation by Estabrook $\&$ Wahlquist \cite{Estabrook75} for the GW detection using Doppler spacecraft tracking, for which the principle is similar to pulsar timing, Detweiler explicitly showed that ``\textit{a gravitational wave incident upon either a pulsar or the Earth changes the measured frequency and appears then as a anomalous residual in the pulse arrival time}" \cite{Detweiler1979}. In 1983, Hellings $\&$ Downs \cite{Hellings_Downs83} used timing data from four pulsars to constrain the energy density of any stochastic background to be $\lesssim10^{-4}$ times the critical cosmological density at $\lesssim10^{-8}$ Hz (see, refs. \cite{RomaniTaylor83,Bertotti83} for similar results). Almost simultaneously to these works, Backer et al. discovered the first millisecond pulsar PSR B1937+21 \cite{Backer82MSP}. Because of its far better rotational stability than any previously known pulsars and relatively narrower pulses, timing observations in the following years improved the limit on stochastic backgrounds very quickly and by orders of magnitude \cite{Davis85,Rawley87,Stineb90,Kaspi94}.

Measurements with a single pulsar can not make definite detections of GWs whose effects may be indistinguishable from other noise processes such as irregular spinning of the star itself. By continued timing of an array of millisecond pulsars, i.e., by constructing a pulsar timing array (PTA), GWs can be searched for as correlated signals in the timing data. Hellings $\&$ Downs \cite{Hellings_Downs83} first did such a correlation analysis to put limits on stochastic backgrounds. Romani \cite{Romani89} and Foster $\&$ Backer \cite{Foster_Backer90} explored the greater scientific potential of PTA experiments: (1) searching for GWs; (2) providing a time standard for long time scales; and (3) detecting errors in the solar system ephemerides.

\begin{figure}[H]
\centering
\includegraphics[width=0.45\textwidth]{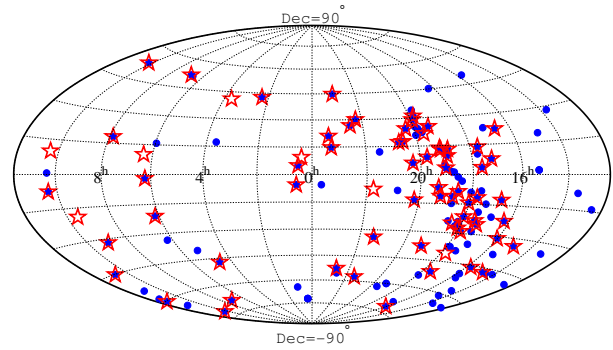}
\caption{Distribution of millisecond pulsars on the sky in equatorial coordinates, containing 70 IPTA pulsars (red stars) and all presently known pulsars that have pulse periods $P<15$ ms and $\dot{P}<10^{-19}\,{\rm{s}}{\rm{s}}^{-1}$ (blue dots; as found in the ATNF Pulsar Catalogue version 1.53 \cite{ATNF05Pulsar})}.
\label{fig:SkyMSP}
\end{figure}

There are three major PTAs currently operating: the Parkes PTA (PPTA\footnote{\url{http://www.atnf.csiro.au/research/pulsar/ppta/}}; \cite{PPTA2013,PPTA13CQG}) was set up in 2004 using the Parkes radio telescope in Australia; the European PTA (EPTA\footnote{\url{http://www.epta.eu.org/}}; \cite{EPTA}) was initiated in 2004/2005 and makes use of telescopes in France, Germany, Italy, the Netherlands and the UK; and the North American Nanohertz Observatory for Gravitational Waves (NANOGrav\footnote{\url{http://nanograv.org/}}; \cite{NANOGrav}) was formed in 2007 and carries out observations with the Arecibo and Green Bank telescopes. We summarize in Table \ref{tb:PTAinfo} information about the telescopes used and the number of pulsars monitored by these PTAs. Recently the three PTA collaborations were combined to form the International PTA (IPTA\footnote{\url{http://www.ipta4gw.org/}}; \cite{IPTA,IPTAdick13,Maura_IPTA14}). Looking into the future, GW detection with pulsar timing observations is one of major science goals for some powerful future radio telescopes, such as the planned Xinjiang Qitai 110 m radio telescope (QTT; which is a fully-steerable single-dish telescope \cite{QTT}) and the Five-hundred-meter Aperture Spherical Telescope (FAST; which is expected to come online in 2016 \cite{FAST11,GeorgeFAST14}) in China, and the planned Square Kilometer Array (SKA) and its pathfinders (see ref. \cite{Lazio13CQG} and references therein). Figure \ref{fig:SkyMSP} shows the distribution of IPTA pulsars on the sky along with those presently known millisecond pulsars that may be useful for PTA research with future telescopes.

\begin{table}[H]
\begin{center}
\caption{Information about pulsar timing array projects.}\label{tb:PTAinfo}
\vspace{2mm}
\footnotesize
\begin{tabular*}{0.39\paperwidth}{lcccc}
\hline
     &  \textit{Telescope} & \textit{Diameter} (m)  & \textit{Country}   & \textit{ pulsars}$^{\rm{a}}$ \\
\hline
  PPTA &  Parkes & 64 & Australia   & 20 \\
\hline
  \multirow{5}{*}{EPTA} &  Effelsberg & 100 & Germany   & \multirow{5}{*}{22} \\
   &  Lovell & 76.2 & UK  &   \\
   &  Nancay & $94^{\rm{b}}$ & France  &    \\
   &  Sardinia & 64 & Italy  &    \\
   &  Westerbork & $96^{\rm{b}}$ & Netherlands  &    \\
\hline
  \multirow{2}{*}{NANOGrav} & Arecibo & 305 & \multirow{2}{*}{USA}  & \multirow{2}{*}{22} \\
   & GBT & 100 &   &  \\
\hline
\end{tabular*}
\end{center}
\vspace{-2mm}
\footnotesize
\emph{Notes:} $^{\rm{a}}$ We present the number of pulsars that have been timed for more than 5 years for each project here (see table 3 in ref. \cite{IPTAdick13} for more information). It is worth pointing out that a number of pulsars were recently added to the timing arrays and some pulsars were dropped from these arrays. The total number of pulsars that are currently being observed for IPTA is 70, of which 13 are timed by two timing arrays and 10 by all three arrays. \newline $^{\rm{b}}$ Values of circular-dish equivalent diameter \cite{EPTA10}.
\end{table}

In this paper, we first introduce the basics of pulsars and pulsar timing techniques in section \ref{sec:psr} and \ref{sec:pt} respectively. In section \ref{sec:noise} we highlight some major noise sources that affect PTA's sensitivities to GWs. In section \ref{sec:GWpta} we describe how a PTA responds to GWs. Section \ref{srcPTA} contains a review on GW sources and related results derived from the latest analysis of PTA data. Finally we discuss future prospects in section \ref{sec:conclu}.
\section{Pulsars}
\label{sec:psr}
Neutron stars are born as compact remnants of core collapse supernovae during the death of massive main sequence stars with masses of around $8$--$25\, M_{\odot}$. A pulsar is a highly magnetized, spinning neutron star. It can be detected as it emits beams of electromagnetic radiation along its magnetic axis that is misaligned with its rotational axis. Such beams of radiation sweep over the Earth in the same way lighthouse beams sweep across an observer, leading to pulses of radiation received at the observatory. Because of their exceptional rotational stability, pulsars are powerful probes for a wide range of astrophysical phenomena. As mentioned above, long-term timing observations of PSR B1913+16 provided the first observational evidence of the existence of GWs. Timing observations of PSR B1257+12 led to the first confirmed discovery of planets outside our solar system \cite{ExoplantPT92}. The double pulsar system PSR J0737$-$3039A/B, with both neutron stars having been detected as radio pulsars \cite{Burgay03Nat,Lyne04Sci}, enabled very stringent tests of general relativity and alternative theories of gravity in the strong-field regime \cite{Kramer06Sci}.

\begin{figure}[H]
 \centering
 \includegraphics[width=0.45\textwidth]{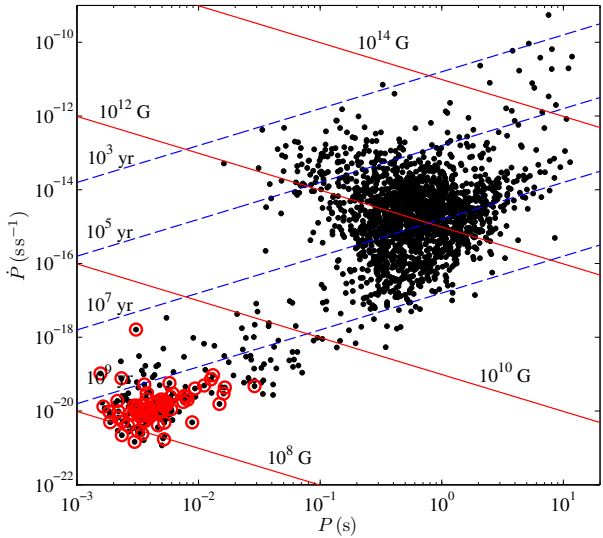}
  \caption{The period ($P$) vs. period derivative ($\dot{P}$) diagram for all known pulsars (black dots; data taken from the ATNF Pulsar Catalogue version 1.53). Red(grey) circles represent the millisecond pulsars currently being timed by the IPTA project. Also shown are lines of constant characteristic age $\tau=P/2\dot{P}$ (dash) assuming that pulsars are spinning down solely because of magnetic dipole radiation \cite{LorimerKramer05}, and of constant inferred surface magnetic field $B_{0}=3.2\times 10^{19}\,\sqrt{P\dot{P}}\,{\rm{Gauss}}$ (solid; \cite{DickTaylor77}). Two distinct populations are apparent in this diagram: (a) \emph{normal pulsars}, with $P\sim$ 0.1--4 seconds and $B_{0}\,\sim10^{11}$--$10^{13}$ Gauss; (b) \emph{millisecond pulsars}, with $P\sim$ 3 milliseconds and $\dot{P}\sim 10^{-20}\,{\rm{s}}{\rm{s}}^{-1}$.}
  \label{fig:PSRppdot}
\end{figure}

The first pulsar was discovered by Bell and Hewish in 1967 \cite{Pulsar68}. Since then over 2500 pulsars have been discovered\footnote{See the ATNF Pulsar Catalogue website \url{http://www.atnf.csiro.au/research/pulsar/psrcat/} for up-to-date information.}, with spin periods ranging from about 1 millisecond to 10 seconds. It is generally believed that pulsars are born with periods of order tens of milliseconds but quickly ($\sim10$ Myr) spin down (because of the loss of rotational energy) to periods of order seconds. This energy loss could be due to a variety of mechanisms, such as the magnetic dipole radiation, emission of relativistic particle winds and even GWs. As pulsars spin down, they eventually reach a point where there is insufficient energy to power electromagnetic radiation. However, for pulsars in binary systems, it is very likely that pulsars are spun up as mass and angular momentum are accreted from their stellar companions. Such an accretion process is observed in X-ray binaries. These pulsars, usually named as \emph{millisecond pulsars}, have spin periods of about several milliseconds and much lower spin-down rates. Figure \ref{fig:PSRppdot} shows the distribution of all known pulsars in the period-period derivative diagram.

\section{Pulsar timing techniques}
\label{sec:pt}
Much of the science based on pulsar observations makes use of the ``pulsar timing" technique \cite{LorimerKramer05,TEMPO2Edwards,LomDemorest13}, which involves measurement and prediction of pulses' times of arrival (TOAs). Individual pulses are generally not useful in this regard as they are unstable and mostly too weak to observe. The average pulse profile over a large number of pulses is stable for a particular pulsar at a given observing wavelength, and therefore very suitable for timing experiments.

The first step in pulsar timing is to measure the \emph{topocentric} pulse arrival times with clocks local to the radio observatories. Data collected with the telescope are \emph{de-dispersed} to correct for frequency-dependent dispersion delays due to the ionized interstellar medium. These data are then \emph{folded} with the period derived from previous observations to form the mean pulse profile. This profile is then correlated with a standard template, either an analytic function or simply a very high signal-to-noise ratio observation, to record the pulse arrival time at the observatory.

The measured TOAs are further transformed to the pulse emission time via a \emph{timing model}, from which the pulse phase of emission is computed. The rotational phase $\phi(t)$ of the pulsar as a function of time $t$ (measured in an inertial reference frame of the pulsar) can be represented as a Taylor series:
\begin{equation}
\phi(t) = \phi(t_{0})+f(t-t_{0})+\frac{1}{2}\dot{f}(t-t_{0})^{2}+\cdots,
\label{PSRphase}
\end{equation}
where $t_0$ is an arbitrary reference time, $f={\rm{d}}\phi/{\rm{d}}t$ is the spin frequency, and $\dot{f}$ is the frequency derivative. A number of corrections are applied when converting the topocentric TOAs to the pulsar frame. Such corrections include:
\begin{enumerate}
\item Clock corrections, which account for differences in the observatory time and a realization of Terrestrial Time (e.g., the International Atomic Time).
\item pulse delay induced by Earth's troposphere.
\item the Einstein delay, i.e., the time dilation due to changes in the gravitational potential of the Earth, the Earth's motion, and the secular motion of the pulsar or that of its binary system.
\item the Roemer delay, i.e., the vacuum light travel time between the observatory and the solar system barycenter, and for pulsars in binaries between the pulsar and the binary system's barycenter.
\item Shapiro delays, i.e., gravitational time delays due to the solar system objects and if applicable the pulsar's companion.
\item Dispersion delays caused by the interstellar medium, the interplanetary medium and the Earth's ionosphere.
\end{enumerate}

The timing model, which describes the above corrections and the pulsar's intrinsic rotational behavior, predicts the rotational phase of the pulsar at any given time as observed from the solar system barycenter. Basic parameters of a pulsar timing model include the spin period, spin-down rate, right ascension and declination of the pulsar, the dispersion measure (discussed later in section \ref{sec:noise}), and Keplerian orbital parameters if the pulsar is in a binary system. Measured TOAs are compared with predictions based on the timing model, and the differences are called \emph{timing residuals}. The (pre-fit) timing residual for the $i$-th observation is calculated as \cite{TEMPO2}:
\begin{equation}
R_{i} = \frac{\phi(t_{i})-N_{i}}{f},
\label{TRsdefi}
\end{equation}
where $N_{i}$ is the nearest integer\footnote{Here note that $\phi(t)$ is measured in \emph{turns} equal to $2\pi$ radians.} to each $\phi(t_{i})$. One can see the key point in pulsar timing is that every single rotation of the pulsar is unambiguously accounted for over long periods (years to decades) of time.

A linear least-squares fitting procedure is carried out to obtain estimates of timing parameters, their uncertainties and the post-fit timing residuals. In practice, this is done iteratively: one starts from a small set of data and only includes the most basic parameters (with values derived from previous observations) so that it is easier to coherently track the rotational phase. Parameter estimates are then improved by minimizing the timing residuals and additional parameters can be included for a longer data set.

The fitting to a timing model and analysis of the timing residuals can be performed with the pulsar timing software package \textsc{TEMPO2} \cite{TEMPO2,TEMPO2Edwards,TEMPO2III}, which is freely available on the internet for download\footnote{www.sf.net/projects/tempo2/}. More recently, an alternative method based on Bayesian inference was also developed \cite{LentatiBayes14,BayesPT14}.

\section{Noise sources in pulsar timing data}
\label{sec:noise}

Timing residuals generally come from two groups of contribution: (1) un-modelled deterministic processes, e.g., an unknown binary companion or a single-source GW; and (2) stochastic processes, e.g., the intrinsic pulsar spin noise and a GW background. Before the discussion of GW detection with PTAs, we briefly discuss some major noise processes in pulsar timing data here.

\subsection*{Radiometer noise}
Radiometer noise arises from the observing system and the radio sky background (including the atmosphere, the cosmic microwave background and synchrotron emission in the Galactic plane). It can be quantified as \cite{LorimerKramer05}:
\begin{equation}
\sigma_{\rm{rad.}} \approx \frac{W}{S/N}\approx \frac{WS_{\rm{sys}}}{S_{\rm{mean}}\sqrt{2\Delta \nu t_{\rm{int}}}}\sqrt{\frac{W}{P-W}}\, ,
\label{Sigma-radio}
\end{equation}
where $W$ and $P$ are the pulse width and period respectively, $S/N$ is the profile signal-to-noise ratio, $S_{\rm{sys}}$ is the system equivalent flux density which depends on the system temperature and the telescope's effective collecting area, $S_{\rm{mean}}$ the pulsar's flux density averaged over its pulse period, $\Delta \nu$ and $t_{\rm{int}}$ are the observation bandwidth and integration time respectively. Radiometer noise can be reduced by using low-noise receivers, observing with larger telescopes, and increasing observing time and bandwidth. One reason to fold individual pulses to obtain an average profile is to reduce the radiometer noise; the reduction is equal to the square root of the number of pulses folded. Radiometer noise is an additive Gaussian white noise and is formally responsible for TOA uncertainties. From equation (\ref{Sigma-radio}), one can see that bright, fast spinning pulsars with narrow pulse profiles allow the highest timing precision.

\subsection*{Pulse jitter noise}
Pulse jitter manifests as the variability in the shape and arrival phase of individual pulses. The mean pulse profile is an average over a large number of single pulses. Although it is stable for most practical purposes, there always exists some degree of stochasticity in the phase and amplitude of the average pulse profile. Pulse jitter noise is intrinsic to the pulsar itself, and thus can only be reduced by increasing observing time, i.e., averaging over more single pulses. Pulse jitter is also a source of white noise, which has been found to be a limiting factor of the timing precision for a few very bright pulsars \cite{Shannon14jitter,1713global}. For future telescopes such as FAST and SKA, jitter noise may dominate over the radiometer noise for many millisecond pulsars \cite{GeorgeFAST14}. Improvement in the timing precision for the brightest pulsar PSR J0437$-$4715 was recently demonstrated with the use of some mitigation methods for pulse jitter noise \cite{Oslow11,Oslow13}.

\subsection*{``Timing noise"}
It has long been realized that timing residuals of many pulsars show structures that are inconsistent with TOA uncertainties \cite{Blandford84,Hob06noise}. Such structures are collectively referred to as \emph{timing noise}. Timing noise is very commonly seen in normal pulsars and has also become more prominent in a number of millisecond pulsars as timing precision increases and the data span grows. The exact astrophysical origins of timing noise are not well understood. It is mostly suggested to be related to rotational irregularities of the pulsar and therefore it is also usually called as spin noise \cite{Hob10noise,RyanS10noise}. For millisecond pulsars, power spectra of timing residuals can typically be modelled as the sum of white noise and red noise. For red noise, a power law spectrum with a low-frequency turnover appears to be a good approximation (see, e.g., figure 11 in ref. \cite{PPTA2013} for analyses of 20 PPTA pulsars).

The presence of red noise results in problems to the pulsar timing analysis as the standard least-squares fitting of a timing model assumes time-independent TOA errors. Blandford et al. \cite{Blandford84} analytically showed the effects of timing noise on the estimates of timing model parameters and suggested the use of the noise covariance matrix to pre-whiten the data for improved parameter estimation. More recently, Coles et al. \cite{Coles2011} developed a whitening method that uses the Cholesky decomposition of the covariance matrix of timing residuals to whiten both the residuals and the timing model. By doing so, noise in the whitened residuals is statistically white and the ordinary least-squares solution of a timing model can be obtained. van Haasteren $\&$ Levin \cite{vHasLevin13} developed a Bayesian framework that is capable of simultaneously estimating timing model parameters and timing noise spectra.

\subsection*{Dispersion measure variations}
Because of dispersion due to the interstellar plasma, pulses at low frequencies arrive later than at high frequencies. Specifically, this dispersion delay is given by:
\begin{equation}
\Delta_{\rm{DM}}\approx (4.15\, {\rm{ms}})\, {\rm{DM}}\, \nu_{\rm{GHz}}^{-2},
\label{Deff}
\end{equation}
where $\nu_{\rm{GHz}}$ is the radio frequency measured in GHz, and dispersion measure (DM, measured in pc cm$^{-3}$) is the integrated column density of free electrons between an observer and a pulsar. Because of the motion of the Earth and the pulsar relative to the interstellar medium, the DM of a pulsar is not a constant in time. Such DM variations introduce time-correlated noise in pulsar timing data.

Pulsars in a timing array are usually observed quasi-simultaneously at two or more different frequencies. For example, the PPTA team observes pulsars at three frequency bands -- 50 cm ($\sim$700 MHz), 20 cm ($\sim$1400 MHz), and 10 cm ($\sim$3100 MHz) -- during each observing session (typically 2-3 days). This makes it possible to account for DM variations and thus reduce the associated noise with various methods \cite{NANOGrav2012,MikeDM2013,KJLee14dm}. The method currently being used by the PPTA is described in Keith et al. \cite{MikeDM2013}, which was built on a previous work by You et al. \cite{YouXP07}. In this method timing residuals are modelled as the combination of a (radio-)wavelength-independent (i.e., common-mode) delay and the dispersion delay. Both components are represented as piece-wise linear functions and can be estimated through a standard least-squares fit. A key feature of this method is that GW signals are preserved in the common-mode component. Ultra wide-band receivers are in development by various collaborations in order to better correct for noise induced by DM variations (along with other benefits such as increasing timing precision, studies of interstellar medium, and etc.).

\subsection*{Interstellar scintillation}
Interstellar scintillation refers to strong scattering of radio waves due to the spatial inhomogeneities in the ionized interstellar medium \cite{Narayan92}, analogous to twinkling of stars due to scattering in the Earth's atmosphere. There are multiple effects associated with interstellar scintillation that cause time-varying delays in measured TOAs, with the dominant one being pulse broadening from multipath scattering \cite{Stinebring13,CordSha10}. Various mitigation techniques have been developed for this type of noise \cite{CordSha10,Demor11Cyclic,LiuK14}. Generally speaking, noise induced by interstellar scintillation is a Gaussian white noise, and can be reduced by increasing observing time and bandwidth. For millisecond pulsars that are observed at current radio frequencies by PTAs, the effects of scintillation are predicted to be small. However, when pulsars are observed at lower frequencies, or more distant (and more scintillated) pulsars are observed, these effects can become more important \cite{CordSha10,Cordes15DM}.

\subsection*{Correlated noise among different pulsars}
The above noise processes are generally thought to be uncorrelated among different pulsars. In a PTA data set that we hope to detect GWs, some correlated noise may be present. For example, (1) instabilities in Terrestrial Time standards affect TOA measurements of all pulsars in exactly the same way, i.e., clock errors result in a \emph{monopole} signature in a PTA data set; (2) the solar system ephemerides, which provide accurate predictions of the masses and positions of all the major solar system objects as a function of time, are used to convert pulse arrival times at the observatory to TOAs referenced at the solar system barycenter. Imperfections in the solar system ephemerides induce a \emph{dipole} correlation in a PTA data set. Indeed, it has been demonstrated that PTAs can be used (1) to search for irregularities in the time standard and thus to establish a pulsar-based timescale \cite{George_Clock12}, and (2) to measure the mass of solar system planets \cite{Champ10}.

\section{Gravitational waves and pulsar timing arrays}
\label{sec:GWpta}
The effects of GWs in a single-pulsar data may be indistinguishable from those due to noise processes as discussed in the previous subsection. Indeed, even without any such noise, GWs that have the same features as those due to uncertainties in the timing model parameters would still be very difficult to detect with only one pulsar. Therefore analysis of single-pulsar data can be best used to constrain the strength of potential GWs \cite{JenetWen04,YiShuXu14MN}.

A PTA is a Galactic-scale GW detector. If one wishes to have an analogy to a laser interferometer, pulsars in the timing array are ``test masses"; pulses of radio waves act as the laser; and the pulsar-Earth baseline is a single ``arm". Millisecond pulsars in our Galaxy, typically $\sim$kpc (or thousands of light years) away, emit radio waves that are received at the telescope with extraordinary stability. A GW passing across the pulsar-Earth baseline perturbs the local spacetime along the path of radio wave propagation, leading to an apparent redshift in the pulse frequency that is proportional to the GW strain amplitude. Let us first consider the special case where a linearly polarized GW propagates in a direction perpendicular to the pulsar-Earth baseline, the resulting timing residual is given by:
\begin{equation}
r(t)=\int_{0}^{L/c}h(t-\frac{L}{c}+\tau){\rm{d}}\tau,
\label{TRlin0}
\end{equation}
where $L$ is the pulsar distance and we adopt the plane wave approximation\footnote{For sources that are close enough ($\lesssim$ 100 Mpc), it may be necessary to consider the curvature of the gravitational wavefront. This, in principle, would allow luminosity distances to GW sources to be measured via a parallax effect \cite{DengXH2011}.}. With the definition of ${\rm{d}}A(t)/{\rm{d}}t=h(t)$, the timing residual takes the following form:
\begin{equation}
r(t)=\Delta A(t)=A(t)-A(t-\frac{L}{c}).
\label{TRptet}
\end{equation}
Here we can see that $A(t)$ results from the GW induced spacetime perturbation incident on the Earth (i.e., the \emph{Earth term}), and $A(t-\frac{L}{c})$ depends on the GW strain at the time of the radio wave emission (i.e., the \emph{pulsar term}). Typical PTA observations have a sampling interval of weeks and span over $\sim$10 yr, implying a sensitive frequency range of $\sim$1--100 nHz. Therefore PTAs are sensitive to GWs with wavelengths of several light years, much smaller than the pulsar-Earth distance.

In the general case where a GW originates from a direction $\hat{\Omega}$, the induced timing residuals can be written as:
\begin{equation}
\label{TR1}
r(t,\hat{\Omega}) = F_{+}(\hat{\Omega})\Delta A_{+}(t) + F_{\times}(\hat{\Omega}) \Delta A_{\times}(t),
\end{equation}
where $F_{+}(\hat{\Omega})$ and $F_{\times}(\hat{\Omega})$ are antenna pattern functions as given by \cite{Wahlq87}:
\begin{eqnarray}
F_{+}(\hat{\Omega}) &=& \frac{1}{4(1-\cos\theta)}\{(1+\sin^2 \delta)\cos^2 \delta_p \cos[2(\alpha-\alpha_p)]\nonumber\\&&\hspace{-14mm} - \sin2\delta \sin2\delta_p\cos(\alpha-\alpha_p) + \cos^2 \delta (2-3\cos^2 \delta_p)\}
\label{Fp}
\end{eqnarray}
\begin{eqnarray}
F_{\times}(\hat{\Omega}) &=& \frac{1}{2(1-\cos\theta)}\{\cos \delta \sin 2\delta_p \sin(\alpha-\alpha_p)\nonumber\\&& - \sin \delta \cos^2 \delta_p \sin[2(\alpha-\alpha_p)]\} ,
\label{Fc}
\end{eqnarray}
where $\cos\theta = \cos\delta \cos\delta_p \cos(\alpha-\alpha_p)+\sin\delta \sin\delta_p$, $\theta$ is the opening angle between the GW source and pulsar with respect to the observer, and $\alpha$ ($\alpha_p$) and $\delta$ ($\delta_p$) are the right ascension and declination of the GW source (pulsar) respectively. The source-dependent functions $\Delta A_{+,\times}(t)$ in equation (\ref{TR1}) are given by:
\begin{equation}
\label{TR2}
\Delta A_{+,\times}(t) = A_{+,\times}(t)-A_{+,\times}(t_p)
\end{equation}
\begin{equation}
\label{TpTe}
t_p = t-(1-\cos\theta)\frac{L}{c}.
\end{equation}
The forms of $A_{+}(t)$ and $A_{\times}(t)$ depend on the type of source that we are looking for.

\subsection*{The Hellings-Downs curve}
\begin{figure}[H]
  \centering
  \includegraphics[width=0.45\textwidth]{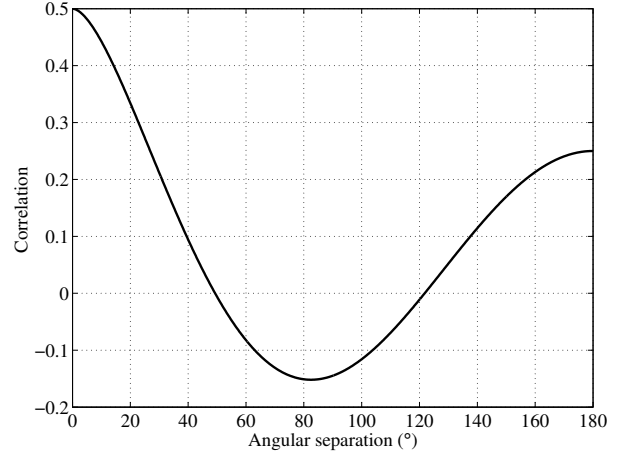}
  \caption{The Hellings-Downs curve, which depicts the expected correlation in timing residuals due to an isotropic stochastic background, as a function of the angular separation between pairs of pulsars.}
  \label{fig:HDcurve}
\end{figure}

An isotropic stochastic background will produce a correlated signal in PTA data sets. Such a correlation uniquely depends on the angular separation between pairs of pulsars, as given by \cite{Hellings_Downs83}:
\begin{eqnarray}
\zeta(\theta_{ij}) &=& \frac{3}{2}\frac{(1-\cos\theta_{ij})}{2}\ln\left[\frac{(1-\cos\theta_{ij})}{2}\right]\nonumber\\&& -\frac{1}{4}\frac{(1-\cos\theta_{ij})}{2}+\frac{(1+\delta_{ij})}{2} ,
\label{HDcorr}
\end{eqnarray}
where $\theta_{ij}$ is the angle between pulsars $i$ and $j$, and $\delta_{ij}$ is 1 for $i=j$ and 0 otherwise. Figure \ref{fig:HDcurve} shows the famous Hellings-Downs curve as given by equation (\ref{HDcorr}) -- it is a factor of $3/2$ larger than the original result of ref. \cite{Hellings_Downs83}. This is because $\zeta(\theta_{ij})$ is normalized to $1$ for the autocorrelation of the stochastic background induced timing residuals for a single pulsar. In Figure \ref{fig:HDcurve} the correlation function takes a value of $0.5$ at zero angular separation as the autocorrelation due to pulsar terms is neglected.

\section{Gravitational wave sources and recent observational results}
\label{srcPTA}
Potential signals that could be detectable for PTAs include: (1) stochastic backgrounds. The primary target is that formed by the combined emission from numerous binary supermassive black holes distributed throughout the Universe. Background signals from cosmic strings (e.g., \cite{Sanidas12}) and inflation (e.g., \cite{ZhaoWen13,TongML14}) have also been studied in the context of PTAs; (2) continuous waves, which can be produced by individual nearby binaries; (3) bursts with memory associated with binary black hole mergers; (4) and all other GW bursts. Below we give a brief overview of these sources and summarize some recent astrophysical results.

\subsection*{Stochastic backgrounds}
The stochastic background from the cosmic population of supermassive binary black holes has been the most popular target for PTA efforts. Generally speaking the signal amplitude depends on how frequently these binaries merge in cosmic history and how massive they are. Both of these quantities are poorly constrained observationally. Assuming that all binaries are in circular orbits and evolve through gravitational radiation only, the characteristic amplitude spectrum of this background is given by \cite{Rajago95,Jaffe_Backer03,Wyithe_Loeb03,WenZL09,Sesana13GWB,Ravi2012,McWilliams14}:
\begin{equation}
h_{c}(f)=A_{\rm{yr}}\left(\frac{f}{f_{\rm{yr}}}\right)^{-2/3},
\label{hcGWBsmbbh}
\end{equation}
where $A_{\rm{yr}}$ is the dimensionless amplitude at a reference frequency $f_{\rm{yr}}=1\, {\rm{yr}}^{-1}$. The fraction of the cosmological critical energy density (per logarithmic frequency interval) contained in the GW background is related to the amplitude spectrum through $\Omega_{\rm{GW}}(f)=(2\pi^{2}/3H_{0}^{2})A_{\rm{yr}}^{2}f_{\rm{yr}}^{2}(f/f_{\rm{yr}})^{2/3}$ where $H_0$ is the Hubble constant. Various models predict a similar range of $A_{\rm{yr}}$, most likely to be $\sim 10^{-15}$ (see, e.g., \cite{WenZL09,Sesana13GWB,Ravi2012}). An exception is the recent model in ref. \cite{McWilliams14} whose prediction is two to five times higher. Recent studies that include the effects of environmental coupling and orbital eccentricities indicate a reduced signal at below $\sim 10$ nHz \cite{Enoki07,Sesana13CQG,Ravi14GWB}.

\begin{figure}[H]
  \centering
  \includegraphics[width=0.46\textwidth]{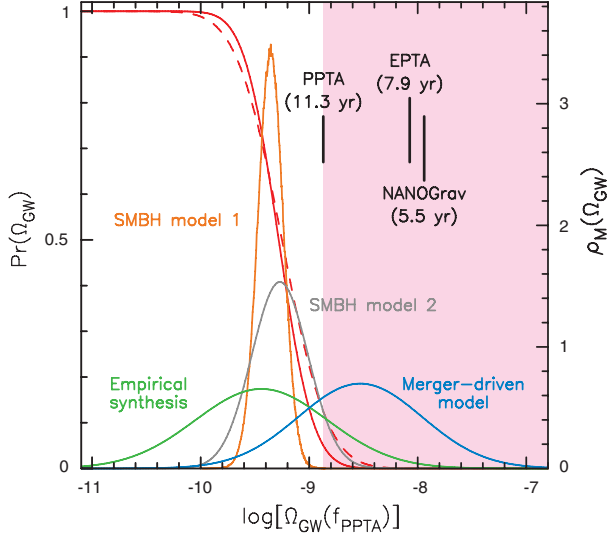}
  \caption{Upper limits on the fractional energy density of the GW background $\Omega_{\rm{GW}}(f)$, at a frequency of 2.8 nHz, as compared against various models of the supermassive binary black hole background. The red(grey) solid and dashed lines show the probabilities ${\rm{Pr}}(\Omega_{\rm{GW}})$ that a GW background with energy density $\Omega_{\rm{GW}}(f)$ exists given the PPTA data, assuming Gaussian and non-Gaussian statistics respectively. Three vertical lines mark the $95\%$ confidence upper limits published by NANOGrav \cite{NANOGrav2012}, EPTA \cite{EPTAlimit} and PPTA \cite{PPTA2013Sci}, where the times next to these lines correspond approximately to the observing spans of the data sets. The shaded region is ruled out with $95\%$ confidence by the PPTA data. Gaussian curves represent the probability density functions $\rho_{M}(\Omega_{\rm{GW}})$ given different models: a merger-driven model for growth of massive black holes in galaxies \cite{McWilliams14}, a synthesis of empirical models \cite{Sesana13GWB}, semi-analytic models (SMBH model 1; \cite{PPTA2013Sci}) based on the Millenium dark matter simulations \cite{Millennium,MillenniumII} and a distinct model for black hole growth (SMBH model 2; \cite{Kulier15}). Figure taken from ref. \cite{PPTA2013Sci}.}
  \label{fig:PPTAlimitGWB}
\end{figure}

Jenet et al. \cite{Jenet05} suggested that it is possible to make a detection of the binary black hole background if 20 pulsars are timed with a precision of $\sim$ 100 ns over $\gtrsim$ 5 years. Three PTAs have searched for such a background signal assuming it is isotropic, leading to increasingly more stringent upper limits on the background strength \cite{Jenet2006,YardleySGWB,EPTAlimit,NANOGrav2012,EPTA15GWB,PPTA2013Sci}. The most constraining limit published to date ($A_{\rm{yr}} <2.4 \times 10^{-15}$) comes from the PPTA collaboration by Shannon et al. \cite{PPTA2013Sci}. As shown in Figure \ref{fig:PPTAlimitGWB}, this limit ruled out the most optimistic model of \cite{McWilliams14} with $90\%$ confidence and is in tension with other models at $\sim50\%$ confidence.

Recently methods have also been proposed to search for a more general anisotropic background signal \cite{Cornish13,Minga13Aniso,Taylor13Aniso,Gair14}. Using the 2015 EPTA data, Taylor et al. \cite{Taylor15Aniso} placed constraints on the angular power spectrum of the background from circular, GW-driven supermassive black hole binaries and found that the data could not update the prior knowledge on the angular distribution of a GW background.

\subsection*{Continuous waves}
Individual supermassive binary black holes, especially the most nearby and/or massive ones, could provide good opportunities for detection of continuous waves. Unlike compact binaries in the audio band, supermassive binary black holes detectable for PTAs are mostly in the early stage of inspiral and therefore emit quasi-monochromatic waves. For an inspiralling circular binary of component masses $m_1$ and $m_2$, the GW strain amplitude is given by \cite{Thorne87}:
\begin{equation}
h_{0}=2\frac{(G M_c)^{5/3}}{c^{4}}\frac{(\pi f)^{2/3}}{d_{L}}
\label{h0},
\end{equation}
where $d_L$ is the luminosity distance of the source, and $M_c$ is the chirp mass defined as $M_c = M \eta ^{5/3}$, with $M= m_1+m_2$ the total mass and $\eta = m_1 m_2 /M^{2}$ the symmetric mass ratio. After averaging over the antenna pattern functions given by equations (\ref{Fp}-\ref{Fc}) and the binary orbital inclination angle, the Earth-term timing residuals induced by a circular binary is $\sim h_{0}/2\pi f$.

Before the establishment of major PTAs, Jenet et al. \cite{JenetWen04} developed a framework in which pulsar timing observations can be used to constrain properties of supermassive binary black holes and applied the method to effectively rule out the claimed binary black hole system in 3C 66B \cite{3C66B03}. In recent years growing efforts have gone into investigating the detection prospects \cite{Sesana2009,KJLee2011,Chiara12PRL,Ravi14Single,Rosado15} of, and designing data analysis methods \cite{Babak2012,Ellis2012,EllisBayesian,Taylor14,YanWang14,PPTAcw14,Zhu15Single} for, continuous waves.

\begin{figure}[H]
  \centering
  \includegraphics[width=0.45\textwidth]{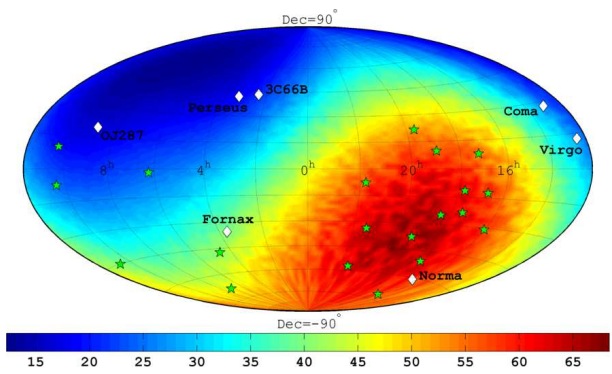}
  \caption{Sky distribution of luminosity distance (in Mpc) out to which a circular binary black hole of chirp mass $10^{9} M_{\odot}$ and orbital period of 5 nHz could be detected with the PPTA data set. Sky locations of the 20 PPTA pulsars are indicated by ``$\star$". White diamonds mark the location of possible supermassive black hole binary candidates or nearby clusters: Virgo (16.5 Mpc), Fornax (19 Mpc), Norma (67.8 Mpc), Perseus (73.6 Mpc), 3C 66B (92 Mpc), Coma (102 Mpc) and OJ287 (1.07 Gpc). Figure originally published in ref. \cite{PPTAcw14}.}
  \label{fig:SkySenDR1}
\end{figure}

Yardley et al. \cite{Yardley2010} calculated the first sensitivity curve of a PTA to this type of sources using an earlier PPTA data set presented in ref. \cite{Verbiest09}. Recently both PPTA \cite{PPTAcw14} and NANOGrav \cite{NANOcw14} conducted searches for continuous waves in their corresponding real data sets. Because of its excellent data quality, PPTA has achieved by far the best sensitivity for continuous waves. Figure \ref{fig:SkySenDR1} shows a sky map of PPTA's sensitivities to circular binary black holes of chirp mass $10^{9} M_{\odot}$ and orbital period of $\sim$6 years \cite{PPTAcw14}. Unfortunately most nearby galaxy clusters or binary black hole candidates are within the less sensitive sky region. Zhu et al. \cite{PPTAcw14} also presented very stringent upper limit\footnote{More recently, the EPTA group obtained slightly better limits for $f<$ 10 nHz and comparable results at higher frequencies \cite{EPTAcw15}.} on the GW strain amplitude ($h_0<1.7\times 10^{-14}$ at 10 nHz) and on the local merger rate density of supermassive binary black holes (fewer than $4 \times 10^{-3} {\rm{Mpc}}^{-3} {\rm{Gyr}}^{-1}$ for $M_{c}\geqslant10^{10} M_{\odot}$).

\subsection*{Gravitational wave memory}
A GW memory is a permanent distortion in the spacetime metric \cite{Memory87,Favata09}. Such effects can be produced during mergers of supermassive binary black holes and cause instantaneous jumps of pulse frequency. For a single pulsar, this is indistinguishable from a glitch event. With a timing array, GW memory effects can be searched for as simultaneous pulse frequency jumps in all pulsars when the burst reaches the Earth. It has been suggested that GW memory signals are in principle detectable with current PTAs for black hole mass of $10^{8}\, M_{\odot}$ within a redshift of 0.1 \cite{Seto09,GWM-RvH,Pshirkov10,Dusty14GWM}. However, the event rate is highly uncertain. Current estimates are very pessimistic, predicting only $0.03$ to $0.2$ detectable events every 10 years for future PTA observations based on the SKA \cite{CordesJenet12,Ravi14Single}. Actual searches in existing PTA data sets -- see ref. \cite{GWM_PPTA} for PPTA and ref. \cite{NANO15Memory} for NANOGrav -- have set upper limits on the memory event rate, which remain orders of magnitude above theoretical expectations.

\subsection*{Gravitational wave bursts}
Potential burst sources of interest to PTAs include the formation or coalescence of supermassive black holes, the periapsis passage of compact objects in highly elliptic or unbound orbits around a supermassive black hole \cite{Finn10ApJ}, cosmic (super)string cusps and kinks \cite{Vilenkin81,Damour00,Siemens06}, and triplets of supermassive black holes \cite{TripleBH10}. Finn $\&$
Lommen \cite{Finn10ApJ} developed a Bayesian framework for detecting and characterizing burst GWs (see also ref. \cite{Deng14PRD}). Zhu et al. \cite{Zhu15Single} recently proposed a general coherent (frequentist) method that can be used to search for GW bursts with PTAs. No specific predictions have been made for detecting GW bursts with PTAs in the literature. There has been no published results on searches for bursts using real PTA data.

\section{Summary and future prospects}
\label{sec:conclu}
With their existence predicted by general relativity one century ago, GWs have not yet been directly detected. However, it is widely believed that we are on the threshold of opening the gravitational window into the Universe. In the audio band (from 10 to several kilo Hz), 2nd-generation laser interferometers such as Advanced LIGO are about to start scientific observations and a detection of signals from compact binary coalescences (e.g., binary neutron star inspirals) is likely within a few years (see the article by Reitze et al. in this issue). In the nanohertz frequency range, PTA experiments have achieved unprecedented sensitivities and started to put serious constraints on the cosmic population of supermassive black hole binaries.

The sensitivity of a PTA can be improved by a) increasing the data span and observing cadence, b) including more pulsars in the array, and c) reducing the noise present in the data. Regarding the first two factors, the combination of data sets from three PTAs to a single IPTA data set offers the most straightforward benefit. Other ongoing efforts include optimization of observing strategies, searches for millisecond pulsars, characterization of various noise processes and corresponding mitigation methods, and development of advanced instrumentations. In the longer term, pulsar timing observations with FAST and SKA will provide advances in all aspects of PTA science, not only leading to the detection of GWs but also allowing detailed studies of the nanohertz gravitational Universe.

\vspace*{2mm} \Acknowledgements{\bahao XZ, LW and GH acknowledge funding support from the Australian Research Council.}

\bibliographystyle{unsrt}
\bibliography{PTArev}

\end{multicols}

\end{document}